\documentclass[12pt,aps,prb]{revtex4}
\usepackage{graphicx}

\begin{document}
\title{Change of the plane of oscillation of a Foucault pendulum from simple pictures}
\author{Thomas F. Jordan}
\email{tjordan@d.umn.edu}
\author{J. Maps}
\affiliation{Department of Physics, University of Minnesota, Duluth, Minnesota 55812}

\begin{abstract}
The change of the plane of oscillation of a Foucault pendulum is calculated without using equations of motion, the Gauss-Bonnet theorem, parallel transport, or assumptions that are difficult to explain.
\end{abstract}

\maketitle

The angle of the change in the direction of the plane of oscillation of a Foucault pendulum as it moves in 24 hours around a circle made by the rotation of the Earth is
\begin{equation}
\label{alpha}
\alpha = 2\pi \sin\theta,
\end{equation}
with $\theta $ the latitude angle (from the equator) of the fixed position of the pendulum on the Earth. We will show a simple way to understand this relation. As in other derivations, we assume the Earth is a sphere with an isotropic distribution of mass, and we neglect the centrifugal force. The change of the plane of oscillation is just a result of the Coriolis force.

All we need to know about the Coriolis force is that it is zero at the equator. There is no change of the plane of oscillation of a pendulum at the equator. Our other input observation is that the role of the Earth's rotation is just to move the pendulum around the Earth's mass, which puts it in a moving reference frame. The result would be the same if the Earth did not rotate and something else moved the pendulum. Then it would be the same for motion around any great circle as around the equator. There would be no change of the plane of oscillation if the pendulum were moved around any great circle.

We consider a pendulum moving around a circle of fixed latitude. This path is not a great circle, but we can approximate it as the union of a large number of short segments of great circles, each running along the circle of fixed latitude for a small increment of longitude $\Delta \phi $. The maximum latitude of each great circle is the latitude of the pendulum. It is reached at the center of the segment the great circle provides. This is shown in Fig.~1.

We calculate the change of the plane of oscillation as the pendulum moves along the path made by these segments. There is no change as the pendulum moves along a segment. There is a change only when the pendulum moves from one segment to the next. Then the angle between the plane of oscillation and the path is changed by the angle between the two segments at the point where they meet. The change of the plane of oscillation as the pendulum moves along the path is the sum of these discrete changes. This part is the same as in derivations that go on to use the Gauss-Bonnet theorem. \cite{WilczekShapere,vonBergmann2} We will use simpler mathematics.

The angle $\Delta \alpha $ between two great circles at a point where they meet is the angle between the directions perpendicular to the planes of the two circles. For two great circles that reach the same maximum latitude at an angle $\theta $ from the equator at points that differ in longitude by a small angle $\Delta \phi $, the angle $\Delta \alpha $ is
\begin{equation}
\label{deltaalpha}
\Delta \alpha = \Delta \phi \sin\theta .
\end{equation}
This is easy to see. The angles $\Delta \phi $ and $\Delta \alpha $ are shown in Fig.~2. Rotation through the angle $\Delta \phi $ around the axis of the Earth takes one great circle to the other. It takes the line through the center of the Earth perpendicular to the plane of one circle to the line through the center of the Earth perpendicular to the plane of the other circle. As a vector of length $r$ on the first line is rotated to the other line, its tip moves through an arc of length $\Delta \phi \, r\sin\theta $, because its distance from the axis of the Earth is $r\sin\theta $ (see Figs.~3 and 4). We can conclude that the angle between the lines is $\Delta \phi \sin\theta $ because the arc is almost a straight line when $\Delta \phi $ is small.

The overall result is that $\Delta \phi \sin\theta $ is the change in the direction of the plane of the Foucault pendulum as it moves through an angle $\Delta \phi $ of longitude at latitude $\theta $ from the equator. Since $\sin\theta $ remains constant, this overall result holds for angles that are not small as well as for those that are. For a complete circle, made in 24 hours, the change in the direction of the plane of oscillation is given by Eq.~(\ref{alpha}).

We can use equations to describe key points further. We take the center of the Earth to be the origin of our coordinate system, write $\mathbf{x}$, $\mathbf{y}$, $\mathbf{z}$ for unit vectors in the $x$, $y$, $z$ directions, let the $z$ axis run north along the axis of the Earth, and choose $x$ and $y$ axes so that the vector perpendicular to the first great circle is
\begin{equation}
\label{vecA}
\mathbf{A} = r \sin \theta \, \mathbf{x} + r \cos \theta \, \mathbf{z}.
\end{equation}
This is shown in Fig.~3. The vector $\mathbf{B}$ perpendicular to the next great circle is obtained by rotating $\mathbf{A}$ through $\Delta \phi $ around the $z$ axis. To first order in $\Delta \phi $, it is
\begin{equation}
\label{vecB}
\mathbf{B} = \mathbf{A} + r \sin \theta \, \Delta \phi \, \mathbf{y}.
\end{equation}
The vectors $\mathbf{A}$ and $\mathbf{B}$ are shown in Fig.~4. Using Eq.~(\ref{vecB}), we can calculate the angle $\Delta \alpha $ between $\mathbf{A}$ and $\mathbf{B}$ from
\begin{equation}
\label{veccross}
r^2 \Delta \alpha = |\mathbf{A} \times \mathbf{B}| 
 = r^2 \sin \theta \, \Delta \phi,
\end{equation}
which gives Eq.~(\ref{deltaalpha}),

At its base, built on a property of the Coriolis force, our derivation is firmly tied to the dynamics. It makes no assumptions beyond those made by the standard description \cite{Taylor} that uses equations of motion. It does not require assumptions that are difficult to explain, as an alternative point of view like parallel transport does. \cite{HartMillerMills}


\begin{figure}[h!]
\includegraphics[width=3.3in]{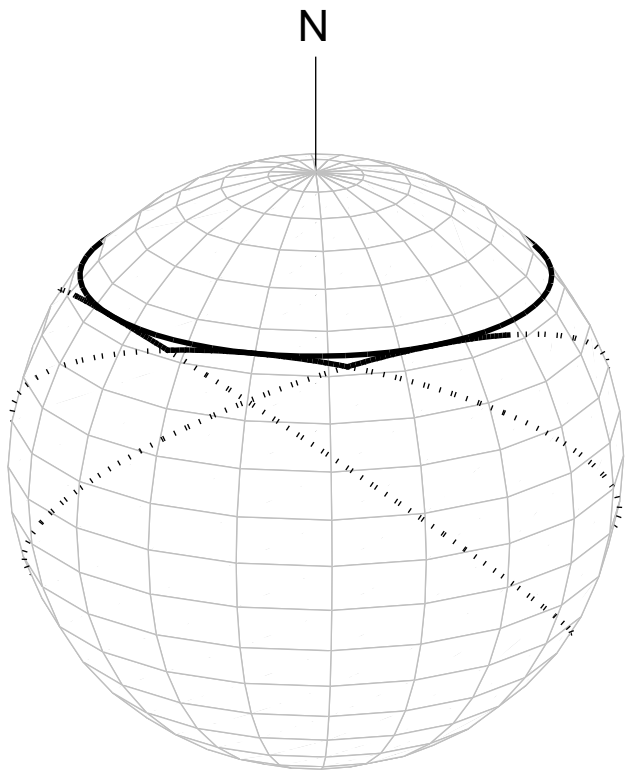} 
\caption{A circle of fixed latitude approximated by segments of great circles.}
\end{figure}

\begin{figure}[h!]
\includegraphics[width=3.3in]{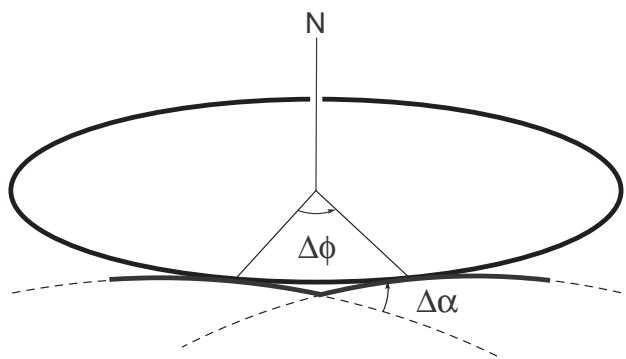}
\caption{The angles $\Delta \phi$ and $\Delta \alpha$. Rotation through $\Delta \phi$ around the axis of the Earth takes one great circle to the next. The angle between the two great circles at the point where they meet is $\Delta \alpha$.}
\end{figure}

\begin{figure}[h!]
\includegraphics[width=3.3in]{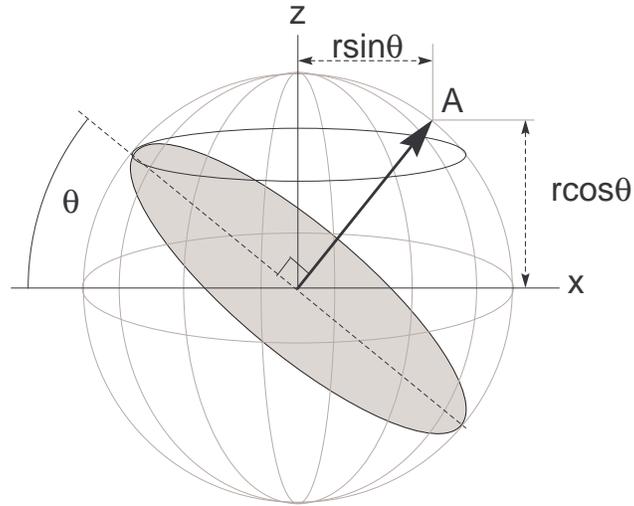} 
\caption{The vector $\mathbf{A}$ perpendicular to one great circle.}
\end{figure}

\begin{figure}[h!]
\includegraphics[width=3.3in]{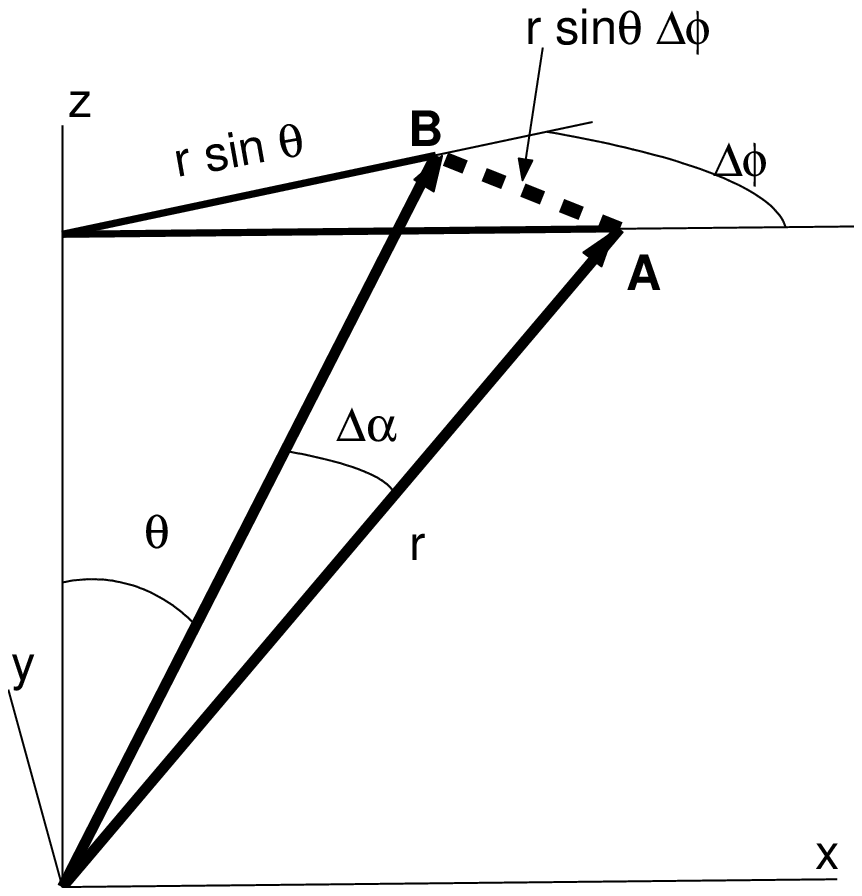}
\caption{
Rotation through $\Delta \phi $ around the $z$ axis takes the vector $\mathbf{A}$ perpendicular to one great circle to the vector $\mathbf{B}$ perpendicular to the next. The tip of the vector moves through the arc shown as a dotted curve. The length of this arc, made by the angle $\Delta \alpha $ between the vectors of length $r$, is  $r\, \Delta \alpha $. For small angles, this length is the same as $r\sin\theta \, \Delta \phi $, the length of the arc made by the angle $\Delta \phi $ between the projections of length $r\sin\theta $ in the $x, y$ plane.
}
\end{figure}


\begin{thebibliography}{4}

\bibitem{WilczekShapere}F. Wilczek and A. Shapere, \textsl{Geometric Phases in Physics} (World Scientific, Singapore, 1989), pp. 3--4. 

\bibitem{vonBergmann2} Jens von Bergmann and HsingChi von Bergmann,
``Foucault pendulum through basic geometry,'' Am. J. Phys. {\bf 75}, 888--892
(2007).

\bibitem{Taylor}J. R. Taylor, \textsl{Classical Mechanics} (University Science Books, Sausalito, 2004). 

\bibitem{HartMillerMills}J. B. Hart, R. E. Miller, and R. L. Mills,
``A simple geometric model for visualizing the motion of a Foucault pendulum,'' Am. J. Phys. {\bf 55}, 67--70
(1987).

\end{thebibliography}
\end{document}